# Shadow operator: Effective dynamic load change operation training in air separation processes based on industrial nonlinear MPC and Bloom's taxonomy

Guanghui Yang[a], Zhijiang Shao[a, b, *], Rui Wang[a], Zuhua Xu[a, b], Lidan Cui[c]

[a] *State Key Lab of Industrial Control Technology, College of Control Science and Engineering, Zhejiang University, Hangzhou 310027, China*
[b] *Alibaba–Zhejiang University Joint Research Institute of Frontier Technologies, Alibaba Group, Hangzhou 310027, China*
[c] *Department of Psychology and Behavioral Sciences, Zhejiang University, Hangzhou 310027, China*

**Abstract**

A novel human–machine interactive training method for dynamic load change operation in air separation processes (ASPs) is proposed. A shadow operator (SO) is developed in this method to train ASP operators through industrial model predictive control (IMPC) and Bloom's taxonomy. First, a nonlinear two-layer IMPC machine algorithm is developed for dynamic load change operation. The IMPC uses a linear parameter varying prediction model and an iterative multi-step linearization algorithm to compute accurate control decisions. Second, a hierarchical human–machine cooperation model is established to improve the effectiveness of operation training. The model is inspired by an educational psychology framework (Bloom's taxonomy) and assists ASP operators in enhancing their dynamic operational skills. Finally, five dynamic training modes of the SO are designed based on the IMPC algorithm and the human–machine cooperation model. The practical application results demonstrate that the SO improves the effectiveness of skill acquisition for novice operators and the safety of dynamic operations.

## 1. Introduction

In the modern process industry, the role of operators has become more demanding than ever before. The demand for advanced machine algorithms in process plants, such as real-time optimization and multivariate process control, has increased significantly over the past three decades [1]. The machine algorithms introduce additional complexity into the process, increasing the possibility of errors and necessitating greater operator proficiency [2]. Nonetheless, research indicates that human error is now one of the leading causes of process accidents [3]. Specifically, misperceptions of process states by operators, misunderstandings of machine algorithm behavior, and improper operation frequently exacerbate minor issues into catastrophic ones [2]. Accidents were found to involve four types of human error: no action (40%), delayed action (31%), aggravated action (24%), and inefficient action (5%) [4]. The lack of effective training methods is the leading cause of human error [2].

In order to improve operator skill levels and operational safety, operator training simulators (OTSs) have been widely deployed in central control rooms (CCRs). Ahmad et al. developed an OTS for homogeneously catalytic two-step biodiesel production using a first-principle model [5]. Using the OTS, they simulated the correct and incorrect behavior of novice operators in a variety of scenarios to hone the operators' capacity to conduct safe operations. Yang et al. developed an OTS using a data-driven model to account for the complexity of dynamic operations in air separation units [6]. The OTS is used to train operators on dynamic operational skills. Sankar et al. discovered that well-trained and competent operators are essential for the safe operation of nuclear power plants. They built an OTS for a 500 MW fast-neutron breeder reactor [7]. Dudley et al. and Park et al. also developed OTSs for training operators of different nuclear power plants [8,9]. In [10], Patle et al. provide a comprehensive summary of OTS applications in the process industry and a discussion of the essential elements of OTS development.

The deployed OTS provides a platform for CCR operators to acquire and enhance operational skills. Typically, the OTS developer's job is completed when the OTS is deployed and can simulate the process with reasonable accuracy [10]. However, efficient operation training methods are often overlooked by OTS developers. In the current training method, skilled operators are typically instructors who provide trainees with operational guidance during operation training [11]. Trainees adjust their operations based on the guidance and feedback from the skill evaluation system [11]. This training method is tacitly accepted and widely adopted in process plants, but the following evidence suggests that it is ineffective and has not kept pace with the development of advanced machine algorithms [2]. First, novice operators spend the majority of their time on trial and error due to the subjectivity and limited availability of expert instructors [11], which reduces the efficiency of training and causes the skill learning curve to rise slowly. Second, as an indispensable element between the operator and physical unit [12], advanced machine algorithms (such as advanced control and optimization algorithms) are entirely ignored in the current training method. Advanced machine algorithms are deeply involved in the daily operations of process plants, and the role of process operators has shifted from physical operation to supervision [12]. In addition to cultivating proficient manual skills, process operators must supervise the automated operation of machine algorithms to intervene quickly with a high level of situational awareness and professional manual skills when algorithmic decisions are abnormal [12]. Process operators must comprehend the operational behavior of machine algorithms and learn to cooperate with or even compete with them in emergencies. Unfortunately, accident investigations have shown that process operators' misunderstanding, distrust, or overtrust in the decision-making results of machine algorithms frequently escalates minor problems into catastrophic accidents [3]. Accidents in other fields, such as the Boeing 737 Max accident [13] and the Tesla accident [14], also indicate that serious disasters will occur if automated machine algorithms and the resulting human–machine interaction issues are neglected during the training phase. Consequently, innovative operator training methods that consider the efficiency of skill acquisition and incorporate advanced machine algorithms are essential for process operators.

This study proposes a novel human–machine interactive operation training method for air separation process (ASP) operators. A shadow operator (SO) is developed as an experienced virtual instructor to train ASP operators during dynamic load change (DLC) operations. First, a nonlinear two-layer industrial model predictive control (IMPC) machine algorithm is developed to take accurate actions during DLC operations. A linear parameter varying (LPV) model is identified as the prediction model, and a multi-step linearization algorithm calculates the control action in the IMPC. Second, to improve the effectiveness of operation training, this study establishes a hierarchical human–machine cooperation (HMC) model. The model is inspired by a cognitive framework in educational psychology (Bloom's taxonomy) and helps operators improve dynamic operational skills. Based on the IMPC algorithm and HMC model, five dynamic operation training modes (identities) of the SO are finally designed. The SO is used to train novice ASP operators. The applications demonstrate that the SO accelerates the learning curve for novice operators and improves the safety of dynamic operations.

This paper is structured as follows. Section 2 describes the characteristics of the ASP and presents a serious accident caused by a machine algorithm. In Section 3, the SO and its five dynamic training modes (identities) are developed. Section 4 discusses the industrial application of the SO in DLC operation training. Finally, the conclusion is presented in Section 5.

## 2. Problem statement

### 2.1. Air separation process

ASPs are designed to produce high-purity oxygen, nitrogen, and argon for downstream industries such as chemicals, healthcare, and steel [15]. ASPs operate at extremely low temperatures (−170°C to −190°C) to separate oxygen, nitrogen, and argon based on their boiling points. In this study, an externally compressed ASP is considered, as shown in Fig. 1. The ASP is located at the Gas Supply Company of Nanjing Iron Steel United Co., Ltd. in China and has a nominal capacity of 20,000 $Nm^3$/h for gaseous oxygen. The ASP is equipped with the JX300XP distributed control system (DCS) developed by Zhejiang SUPCON Technology Co., Ltd. in China.

In Fig. 1, the air is first filtered by the air filter (AF), then compressed by the main air compressor (MAC), cooled by the air cooling tower (ACT), and finally absorbed by the molecular sieve absorber (MSA) to obtain clean air. After heat transfer in the main heat exchanger (MHE), a portion of the purified air is fed to the high-pressure column (HPC). The remaining portion is supplied to the low-pressure column (LPC) through the booster, MHE, and expander.

The LPC and HPC distill air into nitrogen and oxygen products. At the top of the HPC, a portion of the material is separated into the pure liquid nitrogen (LIN) product, a portion is sent to the LPC, and the remainder is returned to the HPC as reflux. At the bottom of the HPC, a portion of the oxygen-enriched air stream is sent to the LPC for further distillation, while the other portion is used as a coolant in the crude argon column II (CAC-II). At the bottom of the LPC, a portion of the liquid oxygen (LOX) stream is stored as the LOX product, while the remaining portion passes through the MHE and becomes the gas oxygen (GOX) product. At the top of the LPC, a portion of the high-purity nitrogen becomes the gas nitrogen (GAN) product after passing through the subcooler and MHE, while another portion containing more oxygen impurities becomes the waste nitrogen (WN).

The CAC-1, CAC-II, and pure argon column (PAC) separate the argon product from the argon-enriched substances. In CAC-I, the argon-enriched stream from the LPC is used as feedstock, while the oxygen-enriched liquid obtained at the bottom is returned to the LPC, and the crude argon obtained at the top is sent to CAC-II. In CAC-II, the bottom stream is returned to CAC-I as reflux, whereas the top stream is fed to PAC as feedstock. In PAC, the argon-enriched material is further purified through distillation, yielding the pure liquid argon (LAR) product at the bottom. After heat exchange in the MHE, a portion of the LAR transforms into the argon gas (GAR) product.

The ASP provides the GOX product to downstream steel and iron manufacturing units. As the downstream demand for the GOX load changes, DLC operations must be performed during the ASP production. Otherwise, it will affect downstream production or lead to GOX dissipation [16]. DLC operations are difficult for ASP operators because they involve multiple variables, nonlinearities, and high operational intensity. (1) Fig. 1 shows substantial energy and material couplings in the ASP, such as multiple streams exchange heat in the MHE. This complexity makes DLC operations challenging. (2) High-purity product specifications, such as the oxygen content of the GOX and LOX products must exceed 99.6%, introduce nonlinearity to DLC operations. (3) To meet downstream demand, a 5% GOX DLC task must be completed within 20 min (the GOX load changes from 19,000 to 20,000 $Nm^3$/h). According to DCS logs, an ASP operator typically performs between 80 and 100 actions during the task, which is high-intensity. Incorrect operations can lead to nitrogen blockage accidents [6], resulting in product and energy wastage and even plant shutdown.

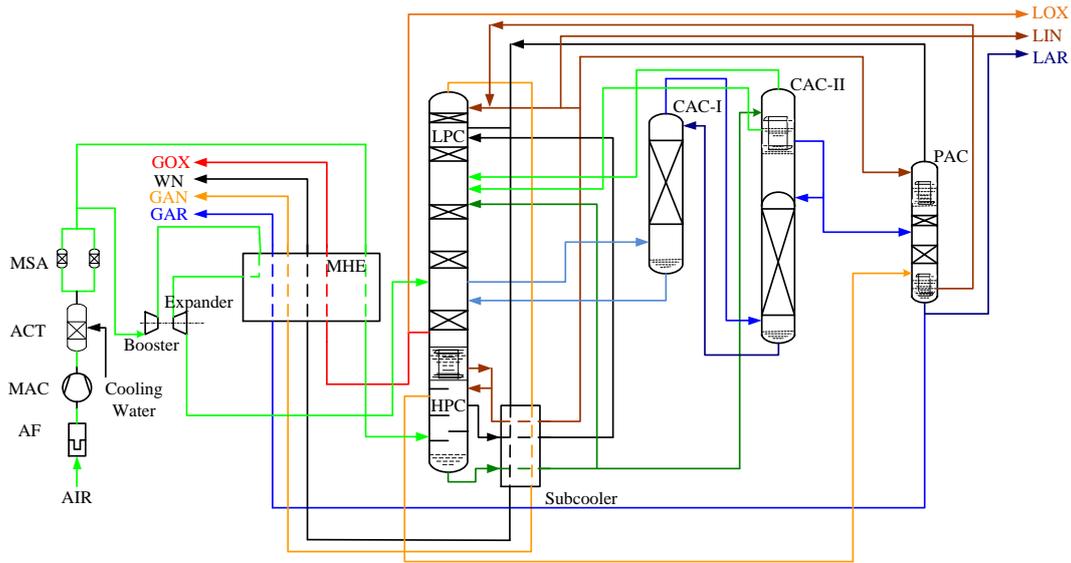

**Fig. 1.** Externally compressed ASP.

Previous work has deployed an OTS and skill assessment system for the DLC operation training in the CCR [6]. During daily training, ASP operators adjust their operations based on the guidance of more experienced operators and the feedback from the assessment system. However, a serious nitrogen blockage production accident (see Section 2.2) caused by a machine algorithm cautioned that the deployed OTS with only conventional process simulation and training methods without considering machine algorithms are far from enough to improve the safety of process operation.

### 2.2. A serious nitrogen blockage accident caused by a machine algorithm

The physical ASP has been equipped with a machine algorithm that automatically performs DLC operations to reduce the operational burden on ASP operators and improve economic efficiency. The machine algorithm is implemented based on IMPC. In daily work practice, the ASP operator supervises the automatic operation of the machine algorithm. Although the machine algorithm performed admirably most of the time, it once caused a serious nitrogen blockage accident due to wrong decision-making results, as shown in Fig. 2.

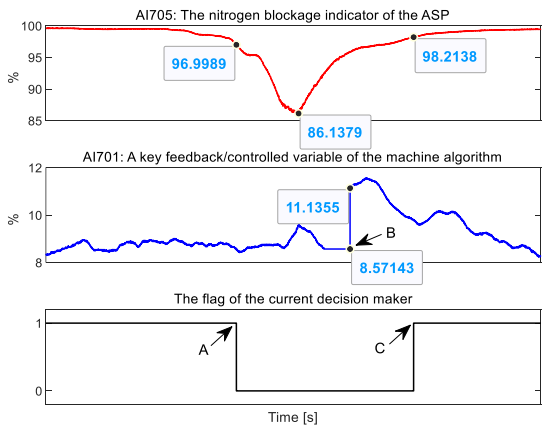

**Fig. 2.** Serious nitrogen blockage accident.

In Fig. 2, the red line is the nitrogen blockage indicator (see AI705 in Table 2). If the indicator falls below the first safety value (97%), a minor nitrogen blockage accident occurs, and the DCS alarm system is activated. If the indicator falls below the second safety value (92%), a serious accident occurs, and the ASP production must be stopped immediately. The blue line is an important feedback/controlled variable of the machine algorithm (see AI701 in Table 2), which has a high weight in the decision-making of the machine algorithm. The black line indicates whether the current decision maker is the human operator or the machine algorithm, and it has only two values, 1 and 0. 1 means the machine algorithm is operating, and 0 means the human operator takes over the operation. This nitrogen blockage accident occurred during the automated DLC operation of the machine algorithm, and three factors contributed to and exacerbated it.

(1) The blue line indicates that an instrumentation engineer manually calibrated the measured value of the key feedback variable at time B. The measured value of this variable was 2.5641% lower than its actual value. Therefore, in conjunction with the red line, it is possible to conclude that the machine algorithm received incorrect feedback during the decision-making process and, as a result, performed the incorrect operation, which led directly to this accident.

(2) The red and black lines indicate that the on-duty operators did not comprehend the abnormal operating action of the machine algorithm in the early stage of the accident. The machine algorithm was not terminated until the nitrogen blockage indicator (97%) triggered the DCS alarm system at time A.

(3) The red and black lines indicate that the operator's manual operation was sluggish and inefficient after turning off the machine algorithm, which caused the nitrogen blockage indicator to fall below the second safety value (92%), and exacerbated the accident. After a long period of accident handling, the ASP operating conditions are stable at time C, and the operator transfers the control authority to the machine algorithm.

The accident investigation revealed that during the 6-h accident, a large amount of argon product was discharged, and the oxygen product purity was extremely unqualified, seriously affecting downstream steel production. On the one hand, the accident shows that the machine algorithm needs to be improved to operate fault-tolerantly when actual measurements are biased. On the other hand, the unanticipated accident serves as a cautionary incident that, in addition to cultivating proficient manual DLC operational skills, a qualified ASP operator must also be familiar with the behavior of the physical machine algorithm to shut it down in the event of wrong decision and then take over the operation with proficient skills.

In summary, developing an effective DLC operation training method that incorporates the machine algorithm for the ASP operators is crucial.

## 3. Development of a SO for effective DLC training

This study proposes a novel human–machine interactive operation training method in which a SO is developed. The SO is expected to be an experienced virtual educator that interacts with and trains ASP operators during DLC operations. This expectation is based on two foundations. (1) The SO can perform the DLC training tasks with excellence. (2) The SO has multiple modes (identities) that can engage in various interactive training activities with the ASP operators. The nonlinear IMPC machine algorithm in Section 3.1 guarantees the first foundation. The HMC model and the five SO identities introduced in Section 3.2 guarantee the second foundation.

### 3.1. Nonlinear IMPC machine algorithm

Considering the complexity of D $b$ LC operations, a nonlinear IMPC machine algorithm is developed in this study. The IMPC employs an LPV prediction model and has a two-layer structure, where the upper layer is a steady-state optimization (SSO) module, and the lower layer is a nonlinear dynamic predictive control (NDPC) module. When the GOX load changes, the SSO module provides the NDPC module with optimal steady-state targets based on an economic indicator. The NDPC module then smoothly and rapidly steers the ASP to the steady-state targets while meeting all controlled variable (CV) and manipulated variable (MV) constraints.

Based on an identified LPV prediction model, SSO, and NDPC, the IMPC machine algorithm within the SO exhibits excellent operational behavior during DLC operations.

#### 3.1.1. LPV prediction model in the IMPC

A prediction model that can capture the dynamic and nonlinear characteristics of the ASP is the base of the IMPC. Although the considered ASP exhibits nonlinear characteristics in the DLC task, it is operated in a relatively orderly rather than chaotic manner [6]. As shown in Fig. 3, the DLC task of the ASP is to adjust the GOX load between four steady-state working points: 18,000, 19,000, 20,000, and 21,000 Nm³/h. Note that the GOX load can be adjusted between any two working points.

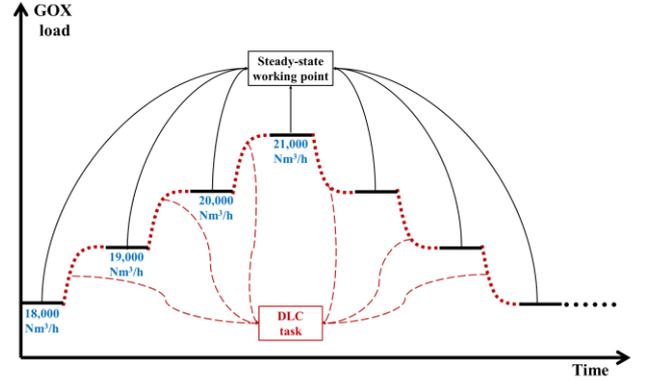

**Fig. 3.** DLC tasks of the ASP.

In this study, an LPV model [6,17,18] is established to capture such nonlinear dynamic characteristics in the DLC task. The identified LPV model has a weighted model structure in which local linear models at the four working points are first identified, and then these local models are nonlinearly weighted to obtain the LPV model.

Consider a multi-input multi-output (MIMO) LPV model with a weighted model structure, and the $k$th output is written as follows:

$$y_k(t) = \sum_{j=1}^{n_w} \alpha_k^j(w) y_k^j(t) + v_k(t) \quad (1)$$

$$k = 1, \cdots n_y$$

where $n_w$ is the number of working points, $y_k^1(t), y_k^2(t), \cdots, y_k^{n_w}(t)$ are the local outputs at the $n_w$ working points, $\alpha_k^1(w), \alpha_k^2(w), \cdots, \alpha_k^{n_w}(w)$ are the nonlinear weighting functions of the working point variable $w$, $v_k(t)$ is the output disturbance, and $n_y$ is the number of model outputs. Note that in this study, the GOX load is the working point variable because it determines the primary nonlinearity of the ASP [6]. $w$ satisfies the following condition:

$$w_{\min} \leq w_1 < w_2 < \cdots < w_{n_w} \leq w_{\max} \quad (2)$$

where $w_{\min}$ and $w_{\max}$ are the low and high limits of the GOX load, respectively.

Denote the $n_u$ inputs as $u_1(t), u_2(t), \cdots, u_{n_u}(t)$, then the local model outputs $y_k^1(t), y_k^2(t), \cdots, y_k^{n_w}(t)$ are given as follows:

$$y_k^1(t) = \sum_{i=1}^{n_u} G_{k,i}^1(q) u_i(t), \text{ for } w = w_1$$

$$y_k^2(t) = \sum_{i=1}^{n_u} G_{k,i}^2(q) u_i(t), \text{ for } w = w_2 \quad (3)$$

$$\vdots$$

$$y_k^{n_w}(t) = \sum_{i=1}^{n_u} G_{k,i}^{n_w}(q) u_i(t), \text{ for } w = n_w$$

where

$$G_{k,i}^{j}(q) = \frac{\left[b_1^{k,i,j}q^{-1} + \cdots + b_n^{k,i,j}q^{-n}\right]q^{-d^{k,i,j}}}{1 + a_1^{k,i,j}q^{-1} + \cdots + a_n^{k,i,j}q^{-n}} \quad (4)$$

$k = 1, \cdots, n_y; i = 1, \cdots, n_u; j = 1, \cdots, n_w$

is the transfer function model from $u_i(t)$ to $y_k(t)$ at $w_j$, $a$ and are parameters of $G_{k,i}^{j}(q)$, $d^{k,i,j}$ and $n$ are the delay and order of $G_{k,i}^{j}(q)$, respectively, and $q^{-1}$ is the unit delay operator.

Combining Eqs. (1) and (3), the LPV model is rewritten as follows:

$$\begin{aligned} y_k(t) &= \sum_{i=1}^{n_u} G_{k,i}(q,w)u_i(t) + v_k(t) \\ &= \alpha_k^1(w)\left[G_{k,1}^1(q)u_1(t) + \cdots + G_{k,n_u}^1(q)u_{n_u}(t)\right] \\ &+ \alpha_k^2(w)\left[G_{k,1}^2(q)u_1(t) + \cdots + G_{k,n_u}^2(q)u_{n_u}(t)\right] \\ &\quad \vdots \\ &+ \alpha_k^{n_w}(w)\left[G_{k,1}^{n_w}(q)u_1(t) + \cdots + G_{k,n_u}^{n_w}(q)u_{n_u}(t)\right] \\ &+ v_k(t) \end{aligned} \quad (5)$$

$k = 1, \cdots n_y$

where the submodel $G_{k,i}(q,w)$ between $y_k(t)$ and $u_i(t)$ is derived as follows:

$$G_{k,i}(q,w) = \sum_{j=1}^{n_w} \alpha_k^j(w) G_{k,i}^j(q) \quad (6)$$

$k = 1, \cdots, n_y; i = 1, \cdots, n_u$

and the LPV model matrix $G_M(q,w)$ can be written as follows:

$$G_M(q,w) = \begin{bmatrix} G_{1,1}(q,w) & G_{1,2}(q,w) & \cdots & G_{1,n_u}(q,w) \\ G_{2,1}(q,w) & G_{2,2}(q,w) & \cdots & G_{2,n_u}(q,w) \\ \vdots & \vdots & \ddots & \vdots \\ G_{n_y,1}(q,w) & G_{n_y,2}(q,w) & \cdots & G_{n_y,n_u}(q,w) \end{bmatrix} \quad (7)$$

Eqs. (5) and (6) indicate that the nonlinear weighted LPV model can be obtained when the nonlinear weighting function $\alpha_k^j(w)$ and the linear model $G_{k,i}^j(q)$ are identified.

The nonlinear identification method in [6] is used to obtain $\alpha_k^j(w)$ and $G_{k,i}^j(q)$. The obtained LPV model has been shown to capture the nonlinear dynamic behavior of the ASP in DLC tasks with high accuracy [6]. The LPV model is used as the prediction model in the IMPC, allowing the IMPC machine algorithm to make accurate control decisions.

### 3.1.2. SSO module

The SSO module provides the optimal steady-state operating points for the NDPC module based on an economic optimization problem when the GOX demand changes. The SSO module serves two distinct roles, determined by the degrees of freedom (DOFs) of the optimization problem. (1) When the DOF does not meet the control requirements of all CVs, the important CVs (such as product purity) must be ensured based on the priority and weight of the CVs [16,19]. (2) When all CVs are within the constraints, and there are remaining DOF, the SSO module is performed to achieve an economic objective [16,19]. Note that the main production cost of the ASP is the electricity consumed during the feed air compression; hence, the economic objective is to reduce electricity consumption while maintaining productivity.

The economic optimization problem in the SSO module is a linear programming (LP) problem defined as follows:

$$\begin{aligned} \left(U_{sso}^*(t), Y_{sso}^*(t)\right) &= \arg\min_{U_{sso}(t), Y_{sso}(t)} J_{SSO}(t) = \\ & B^T U_{sso}(t) + C^T Y_{sso}(t) + Z^T S(t) \end{aligned}$$

s.t.

$$Y_{sso}(t) = K_M(w) U_{sso}(t) + D(t) \quad (8)$$

$u_{i,\min} \leq u_i(t) \leq u_{i,\max}$

$y_{k,\min} - s_k(t) \leq y_k(t) \leq y_{k,\max} + s_k(t)$

$i = 1, \cdots, n_u, k = 1, \cdots, n_y$

where

$$\begin{aligned} B &= \left[b_1(t), \cdots, b_{n_u}(t)\right]^T, C = \left[c_1(t), \cdots, c_{n_y}(t)\right]^T \\ Z &= \left[z_1(t), \cdots, z_{n_y}(t)\right]^T \\ Y(t) &= \left[y_1(t), \cdots, y_{n_y}(t)\right]^T, U(t) = \left[u_1(t), \cdots, u_{n_u}(t)\right]^T \\ D(t) &= \left[d_1(t), \cdots, d_{n_y}(t)\right]^T, S(t) = \left[s_1(t), \cdots, s_{n_y}(t)\right]^T \\ Y_{sso}(t) &= \left[y_{1,sso}(t), \cdots, y_{n_y,sso}(t)\right]^T \\ U_{sso}(t) &= \left[u_{1,sso}(t), \cdots, u_{n_u,sso}(t)\right]^T \end{aligned} \quad (9)$$

In Eqs. (8) and (9), $u$ and $y$ are the MVs and CVs, respectively; $b$ and $c$ are the cost coefficients of $u$ and $y$, respectively; $K_M(w)$ is the gain matrix of the LPV model matrix $G_M(q,w)$ in Eq. (7) and $d$ is the bias; $u_{\min}$ and $u_{\max}$ are the lower and upper limits of $u$, respectively; $y_{\min}$ and $y_{\max}$ are the lower and upper limits of $y$, respectively; $s$ is the slack variable and $z$ is the cost coefficient.

The optimization results ($U_{sso}^*(t)$ and $Y_{sso}^*(t)$) given by the SSO module are the control targets of the MVs and CVs in the NDPC module.

### 3.1.3. NDPC module

Given the tracking targets ($U_{sso}^*(t)$ and $Y_{sso}^*(t)$), the NDPC module calculates the control actions and drives the ASP to the targets without violating any constraints. Traditionally, two methods have been used to calculate the control actions.

(1) Method 1: Nonlinear MPC [20]. A nonlinear MPC can be designed based on the LPV model. In this work, however, the working point variable (the GOX load) of the LPV model is a CV (see Table 2 in Section 4), and a nonconvex optimization problem with nonlinear model constraints must be solved at each time step.

Nonconvex optimization is hampered by numerical convergence issues and needs considerable computation time.

(2) Method 2: Linear MPC [16]. Eq. (5) indicates that a linear MPC with convex quadratic programming (QP) problem can be built by linearizing the nonlinear ASP based on the working point variable at each time step. However, because of the nonlinearity of the ASP, the prediction results of the linearized model over the prediction horizon at the current time step may be inaccurate, affecting the control performance.

In order to balance control efficiency and performance, an iterative multi-step linearization MPC method for computing the control actions is presented in this study. Denote $P$ and $M$ as the prediction and control horizons, respectively. The idea behind the iterative method is first to provide an initial control action in the control horizon, then to replace the nonlinear LPV model in the prediction horizon with $P$ linear models, and finally to update the original control action by solving a QP problem with $P$ linear model constraints. The procedure described above is repeated until the control action converges. Since this iterative method solves the convex QP problem and uses $P$ linear models to predict the future behavior of the ASP, it improves the control efficiency compared to Method 1 and enhances the control performance compared to Method 2. The specific steps of the iterative method are introduced as follows.

Denote the current time step as $t$, the iteration count as *iterCount*, the initial control action over the control horizon as $U_M^{ini}$, and the initial working point variable as $w^{ini}$:

$$iterCount = 0$$
$$U_M^{ini} = \left[U(t-1)^T, \cdots, U(t-1)^T\right]^T \quad (10)$$
$$w^{ini} = w(t-1)$$

where $U(t-1) = \left[u_1(t-1), \cdots, u_{n_u}(t-1)\right]^T$ and $w(t-1)$ are the control action and working point variable at $t-1$, respectively.

Then, recursively predict the CVs and working point variables over the prediction horizon $Y_{PM}(t)$ and $w_{PM}(t)$ with $U_M^{ini}$:

$$Y_{PM}(t+l_y \mid t) = G_M\left(q, w_{PM}(t+l_y-1 \mid t)\right) U_M^{ini}(l_y)$$
$$l_y = 1, \cdots, M \quad (11)$$

and

$$Y_{PM}(t+l_y \mid t) = G_M\left(q, w_{PM}(t+l_y-1 \mid t)\right) U_M^{ini}(M)$$
$$l_y = M+1, \cdots, P \quad (12)$$

and

$$w_{PM}(t \mid t) = w^{ini} \quad (13)$$

In Eq. (11) and (12), $G_M\left(q, w_{PM}(t)\right)$ is the linearized model at the working point $w_{PM}(t)$, and $U_M^{ini}(l_y)$ is the $l_y$th element of $U_M^{ini}$. The control action is constant after $M$ time steps and $U_M^{ini}(l_y) = U_M^{ini}(M), l_y = M+1, \cdots, P$ holds. Note that $Y_{PM}(t)$ is denoted as:

$$Y_{PM}(t) = \left[y_{1,PM}(t), \cdots, y_{n_y,PM}(t)\right]^T \quad (14)$$

and $w_{PM}(t)$ is the predicted work point variable, which is a CV in $Y_{PM}(t)$.

After obtaining the $P$ linearized models and $Y_{PM}(t)$ over the prediction horizon, the optimal control action can be calculated by solving a QP problem, which is defined as follows:

$$U_M^{*,iter}(t) = \arg\min_{U_M(t)} J_{NDPC}(t) =$$
$$\sum_{l=1}^{P} \left\|Y^{ref}(t+l) - Y_{PM}(t+l \mid t)\right\|_{Q_l}^2 + \sum_{l=1}^{P} \left\|E(l)\right\|_{H_l}^2$$
$$+ \sum_{l=0}^{M-1} \left\|\Delta U(t+l)\right\|_{R_l}^2 + \sum_{l=0}^{M-1} \left\|U(t+l) - U_{sso}^*(t)\right\|_{V_l}^2$$

s.t.

Prediction equation constraints : Eqs.(11–13)

$$y_k^{ref}(t+l_y) = y_k(t) + \left(y_{k,sso}^*(t) - y_k(t)\right)\left(1 - e^{-l_y T/\tau_k}\right) \quad (15)$$
$$y_{k,min} - \varepsilon_k(t+l_y) \le y_{k,PM}(t+l_y \mid t) \le y_{k,max} + \varepsilon_k(t+l_y)$$
$$\Delta u_i(t+l_u) = u_i(t+l_u) - u_i(t+l_u-1)$$
$$u_{i,min} \le u_i(t+l_u) \le u_{i,max}$$
$$\Delta u_{i,min} \le \Delta u_i(t+l_u) \le \Delta u_{i,max}$$
$$l_y = 1, \cdots, P, l_u = 0, \cdots, M-1$$
$$k = 1, \cdots, n_y, i = 1, \cdots, n_u$$

where the vectors $Y^{ref}(t)$ and $E(t)$ are given as follows:

$$Y^{ref}(t) = \left[y_1^{ref}(t), \cdots, y_{n_y}^{ref}(t)\right]^T$$
$$E(t) = \left[\varepsilon_1(t), \cdots, \varepsilon_{n_y}(t)\right]^T \quad (16)$$

In Eqs. (15) and (16), $y_k^{ref}(t)$ is the expected reference trajectory of the $k$th CV, allowing $y_k(t)$ to transition quickly and smoothly to its target; $T$ is the sampling time, and $\tau_k$ is the time constant of $y_k^{ref}(t)$; $\varepsilon_k(t)$ is the relax variable of the $k$th CV; $\Delta u_{min}$ and $\Delta u_{max}$ are the lower and upper limits of $\Delta u$, respectively; $U_M(t)$ and $\Delta U_M(t)$ are the trajectories of the MVs and MVs increments, respectively, over $M$, and satisfies the following equation:

$$\Delta U_M(t) = U_M(t) - U_M(t-1)$$
$$U_M(t) = \left[U(t)^T, \cdots, U(t+M-1)^T\right]^T \quad (17)$$
$$\Delta U_M(t) = \left[\Delta U(t)^T, \cdots, \Delta U(t+M-1)^T\right]^T$$

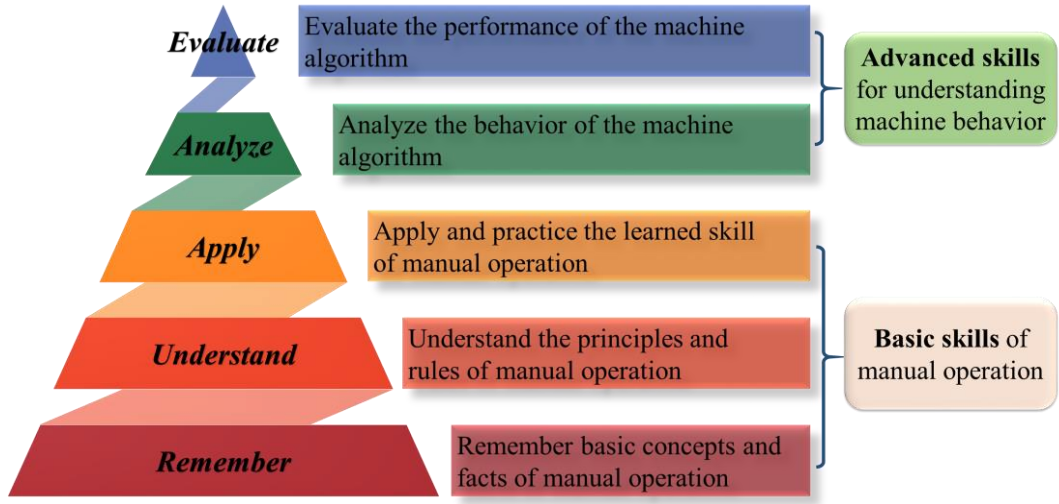

**Fig. 4.** Hierarchical HMC model for effective skill acquisition and improvement of the ASP operators.

The last term of the objective function in Eq. (15) is used to drive the ASP to the desired targets given by the SSO module. The weight matrices $Q_l$, $H_l$, $R_l$, and $V_l$ adjust the relative contribution of the objective function.

Finally, the initial control action $U_M^{ini}$ and iteration count $iterCount$ in Eq. (10) are updated as follows:

$$U_M^{ini} = U_M^{*,iter}(t)$$
$$iterCount = iterCount + 1 \quad (18)$$

Repeat Eqs. (10–12), (15) and (18) until $U_M^{ini}$ converges. Denote the converged control action as $U_M^*(t)$, the first action ($U^*(t)$) of $U_M^*(t)$ is implemented into the ASP. Then the IMPC moves to the next time step.

The developed nonlinear IMPC machine algorithm ensures that the SO executes the DLC tasks excellently (see applications in Section 4). The nonlinear IMPC algorithm is encapsulated as $IMPC(t)$, and its pseudocode is shown in **Algorithm 1**.

---
**Algorithm 1.** The IMPC machine algorithm.
1: **Step 1:** Read the current ASP state on the DCS.
2: **Step 2:** Solve the LP problem of the SSO module:
3: $\quad (U_{sso}^*(t), Y_{sso}^*(t)) = \arg\min_{U_{sso}(t), Y_{sso}(t)} J_{SSO}(t)$.
4: **Step 3:** Update the expected trajectory for each CV:
5:
$$y_k^{ref}(t+l_y) = y_k(t) + (y_{k,sso}^*(t) - y_k(t))(1 - e^{-l_y T/\tau_k})$$
$k = 1, \cdots, n_y, l_y = 1, \cdots, P$
6: **Step 4:** Solve the QP problem of the NDPC module using the iterative multi-step linearization method:
7: $\quad U_M^*(t) = \arg\min_{U_M(t)} J_{NDPC}(t)$.
8: **Step 5:** Implemented $U^*(t)$ into the ASP.

---

### 3.2. Five SO identities for interactive training with ASP operators

Despite possessing exceptional DLC operational abilities, the SO lacks the qualifications to be an instructor. The SO must have different identities to afford the ASP operators various dynamic interactive training modes. In this study, five SO identities are designed. The five SO identities are based on an HMC model and help novice operators acquire basic skills of manual operation and advanced skills to understand machine (the nonlinear IMPC) behavior.

#### 3.2.1. LPV prediction model in the IMPC

The DLC operation training is essentially an educational scenario involving teachers (the SO) instructing students (the ASP operators) to acquire knowledge effectively. Therefore, this study designs the SO's identity for interactive training from an educational psychology perspective.

As a cognitive framework in educational psychology, Bloom's taxonomy categorizes the skills students need to acquire through specialized training and is commonly used as a standard for beginners to master an advanced skill or complicated body of knowledge [21]. Bloom's taxonomy classifies six cognitive levels of knowledge mastery: remember, understand, apply, analyze, evaluate, and create. The six progressive levels represent the scientific steps through which students acquire scientific knowledge [11]. Inspired by Bloom's taxonomy, this study establishes a hierarchical HMC model for effective skill acquisition and improvement of the ASP operators, as presented in Fig. 4. In the HMC model, "human" refers to the ASP operator, while the "machine" refers to the nonlinear IMPC.

Fig. 4 shows the five cognitive levels of knowledge mastery identified by Bloom's taxonomy: remember, understand, apply, analyze, and evaluate. The five cognitive levels are organized in a bottom-to-top approach, representing the easiest to the most challenging to master for novice operators. In the first three cognitive levels, the ASP operator needs to acquire basic manual DLC operational skills, while in the last two levels, the operator needs to acquire advanced skills in understanding the behavior of the machine algorithm (the nonlinear IMPC).

In the five cognitive levels, the SO conducts interactive training with operators in five identities: task performer, cooperation partner, operation advisor, safety supervisor, and troublemaker. Table 1 explains the relationship between the five identities, functions, and cognitive levels.

**Table 1.** Explanation of the five SO identities and functions.

| Levels of training | Identities of the SO | Functions of the SO | Cognitive levels of knowledge mastery |
|---|---|---|---|
| L1 | Task performer | Demonstrate | Remember |
| L2 | Cooperation partner | Assist | Understand and remember |
| L3 | Operation advisor | Advise | Apply and understand |
| L4 | Safety supervisor | Take over | Apply and understand |
| L5 | Troublemaker | Make trouble | Evaluate, analyze, and apply |

Combined with Fig. 4 and Table 1, the five SO identities and functions are explained as follows:

- **L1:** The first identity of the SO is the task performer. The task performer will demonstrate in advance how to operate the DLC task. At this level of training, the ASP operator must remember basic manual operation concepts and facts from a demonstration.
- **L2:** The second identity of the SO is the cooperation partner. The cooperation partner will actively assist and take on some of the MVs as a teammate when the ASP operator performs the DLC task manually. The ASP operator must understand each MV's role and how it changes. At this level of training, the ASP operator is only responsible for part of the MVs, and the cooperation partner operates the rest.
- **L3:** The third identity of the SO is the operation advisor. The operation advisor will advise on the optimal future adjustments for the next few steps for all MVs when the ASP operator needs guidance (issues a help request) during manual operation. Based on the advice, the ASP operator must apply the manual skills learned at the L1 and L2 training levels and further understand the operation principles.
- **L4:** The fourth identity of the SO is the safety supervisor. The safety supervisor will take over the operation when the ASP operator can not continue operating with the current skill level and actively transfers the control authority to the supervisor. The supervisor will not hand over the control authority until the ASP operator issues the command for manual operation. At this level of training, the ASP operator must learn the operation principles when the skill level is insufficient to continue manual operation.
- **L5:** The last identity of the SO is the troublemaker. After training at the four training levels listed above, the ASP operator is considered to have proficient manual skills. This training level simulates the daily working pattern of the ASP operator, where the operator supervises the automatic operation of the machine algorithm (the nonlinear IMPC). At this level, the troublemaker will make intentionally biased or incorrect decisions (troubles) during automatic operations. The ASP operator must evaluate and analyze the behavior of the troublemaker and apply acquired manual skills to resolve such troubles. On the one hand, L5 is challenging for the ASP operators because they need basic manual skills and must be familiar with the machine algorithm's behavior [12]. On the other hand, L5 is critical for the ASP operator because, in industrial practice, the machine algorithm may make unexpected wrong decisions in certain scenarios (see Section 2.2).

So far, the concept of the five SO identities has been proposed from the perspective of educational psychology. The five SO identities provide ASP operators with five dynamic interactive training modes.

### 3.2.2. Implementation algorithm for each SO identity

Based on the proposed five SO identities and the IMPC machine algorithm, this section describes the specific implementation algorithm for each SO identity.

*1. The SO acts as the task performer.*

The function of the task performer is to demonstrate how to operate the DLC task, and its functional diagram is shown in Fig. 5. Note that to improve the training effect at this level, the ASP operator can adjust the ASP state to a previous one by dragging the operation progress bar. Then, the task performer will redemonstrate the operation based on the adjusted process state.

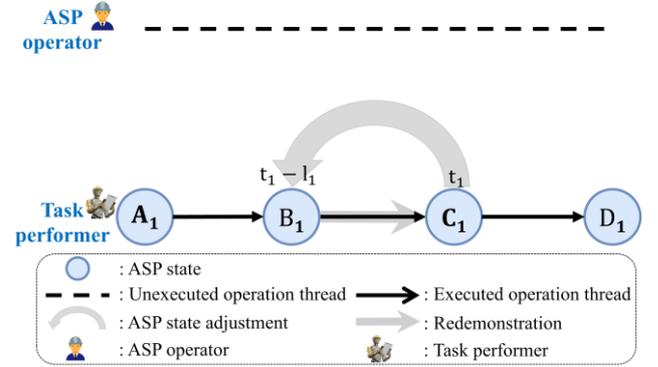

**Fig. 5.** Functional diagram of the task performer.

In Fig. 5, the upper and lower lines represent the operation threads of the ASP operator and the task performer, respectively; the dashed and solid lines indicate the unexecuted and executed threads, respectively. $A_1$ and $D_1$ represent the start and end states of the DLC task, respectively; $C_1$ represents the current state and $B_1$ the previous state that is adjusted back to by the ASP operator. The bold gray arrow $C_1 \to B_1$ represents the ASP state is adjusted from $C_1$ to $B_1$, and $B_1 \to C_1$ represents the redemonstration of the task performer.

Denote the time steps corresponding to states $C_1$ and $B_1$ are $t_1$ and $t_1 - l_1$, respectively, the pseudocode of the task performer is shown in **Algorithm 2**.

**Algorithm 2.** The task performer.

1: **Step 1:** Clear all process states from $t_1 - l_1 + 1$ to $t_1$.
2: **Step 2:** Align the current time step $t$:
3:     $t = t_1 - l_1$.
4: **for** every time step $t = t_1 - l_1, t_1 - l_1 + 1, t_1 - l_1 + 2, \cdots$ **do**
5:   **Step 3:** Call $IMPC(t)$ and calculate $U^*(t)$:
6:     $U^*(t) = IMPC(t)$.
7: **end for**

*2. The SO acts as the cooperation partner.*

The function of the cooperation partner is to cooperate with the ASP operator to complete the DLC task. Unlike the static cooperation method, a dynamic and progressive form of cooperation is provided. The cooperation partner is gradually reducing the number of MVs in charge, while the ASP operator is gradually increasing the number of MVs in charge. At the beginning of the DLC task, the ASP operator is not responsible for any MVs, and the cooperation partner takes on all MVs. As the operation proceeds, the MVs initially in charge of the cooperation partner are gradually transferred to the ASP operator until the operator is responsible for all MVs. Fig. 6 shows the functional diagram of the cooperation partner.

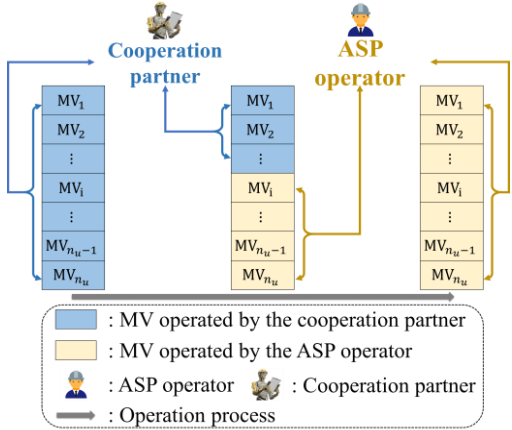

**Fig. 6.** Functional diagram of the cooperation partner.

In Fig. 6, the MVs shaded in pale gold and pale blue indicate that they are operated by the ASP operator and the cooperation partner, respectively. As the operation task proceeds, the ASP operator gradually transitions from operating no MVs to operating all MVs.

Denote $MV^{ASP}$ and $MV^{CP}$ as the MVs operated by the ASP operator and cooperation partner, respectively. Then, the prediction model matrix in Eq. (7) is rewritten as $G_M(q, w, MV^{CP})$:

$$G_M(q, w, MV^{CP}) = \begin{bmatrix} G_{1,1}(q, w, MV^{CP}) & \cdots & G_{1,n_u}(q, w, MV^{CP}) \\ G_{2,1}(q, w, MV^{CP}) & \cdots & G_{2,n_u}(q, w, MV^{CP}) \\ \vdots & \ddots & \vdots \\ G_{n_y,1}(q, w, MV^{CP}) & \cdots & G_{n_y,n_u}(q, w, MV^{CP}) \end{bmatrix} \quad (19)$$

$G_M(q, w, MV^{CP})$ restructures with the MVs operated by the cooperation partner ($MV^{CP}$). Specifically, if the $MV_i$ is not contained in the $MV^{CP}$, the following relationship can be obtained:

$$G_{k,i}(q, w, MV^{CP}) = 0, \quad k = 1, \cdots, n_y \quad (20)$$

Denote the time step corresponding to the current ASP state is $t_2$, the pseudocode of the cooperation partner is shown in **Algorithm 3**.

**Algorithm 3.** The cooperation partner.

1: **for** every time step $t = t_2, t_2 + 1, t_2 + 2, \cdots$ **do**
2:   **Step 1:** Read the current $MV^{CP}(t)$.
3:   **if** $MV^{CP}(t)$ is not equal to $MV^{CP}(t-1)$ **do**
4:     **Step 2:** Update $G_M(q, w, MV^{CP})$.
5:   **end**
6:   **Step 3:** Match the prediction model in Eq. (7):
7:     $G_M(q, w) = G_M(q, w, MV^{CP})$.
8:   **Step 4:** Call $IMPC(t)$ and calculate $U^*(t)$:
9:     $U^*(t) = IMPC(t)$.
10: **end for**

*3. The SO acts as the operation advisor.*

The function of the operation advisor is to provide operational advice (the adjustment values of the MVs) when the ASP operator issues a help request, and its functional diagram is shown in Fig. 7.

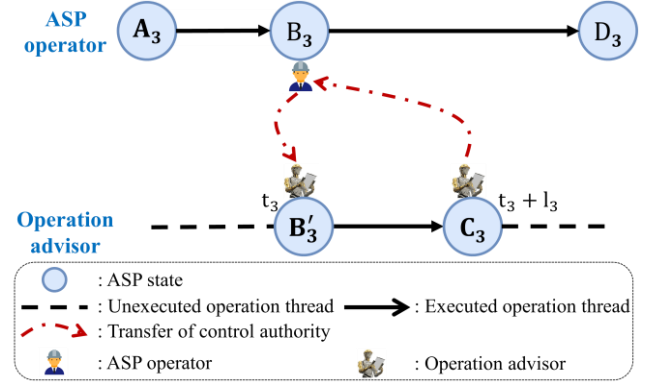

**Fig. 7.** Functional diagram of the operation advisor.

In Fig. 7, the upper and lower lines represent the operation threads of the ASP operator and the operation advisor, respectively; the dashed and solid lines indicate the unexecuted and executed threads, respectively. The red dotted arrow line from the ASP operator icon to the operation advisor icon indicates the transfer of control authority to the operation advisor from the ASP operator and vice versa. $A_3$ and $D_3$ represent the start and end states of the DLC task, respectively; $B_3$ and $B_3^{'}$ are the same state, representing the ASP operator issuing a help request and the operator advisor starting the operation; $C_3$ represents that state that the operator advisor completes its operations.

$B_3^{'} \to C_3$ represents the operation of the advisor, i.e., to deduce the operation of the next few steps. At state $C_3$, the operation advisor shows the deduction result (operational advice) to the ASP operator, and the process state is reset to $B_3$. The ASP operator continues the operation based on the operational advice and state $B_3$.

Denote the time step corresponding to state $B_3$ ($B_3^{'}$) and $C_3$ are $t_3$ and $t_3+l_3$, respectively, the pseudocode of the operation advisor is shown in **Algorithm 4**.

**Algorithm 4.** The operation advisor.
1: **Step 1:** Read and store the process state of $B_3$.
2: **Step 2:** Stop the timer of the ASP operator, and start the timer of the operation advisor.
3: **for** every time step $t = t_3, t_3+1, \cdots, t_3+l_3$ **do**
4:   **Step 3:** Call $IMPC(t)$ and calculate $U^*(t)$:
5:     $U^*(t) = IMPC(t)$.
6: **end for**
7: **Step 4:** Stop the timer of the operation advisor and show the operational advice to the ASP operator.
8: **Step 5:** Reset the ASP state to $B_3$ when the ASP operator requests to continue the operation.
9: **Step 6:** Start the timer of the ASP operator.

*4. The SO acts as the safety supervisor.*

The function of the safety supervisor is to take over the operation when the ASP operator transfers the control authority to the supervisor, and its functional diagram is shown in Fig. 8.

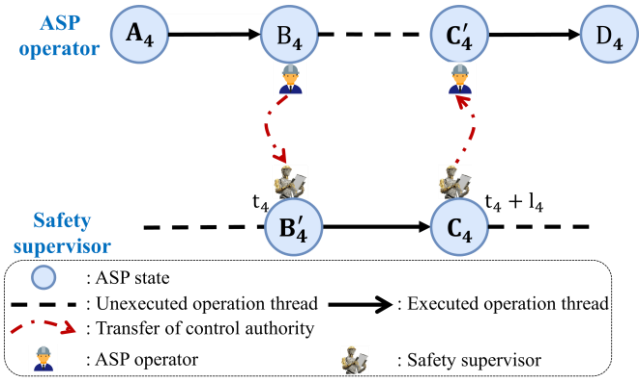

**Fig. 8.** Functional diagram of the safety supervisor.

In Fig. 8, the upper and lower lines represent the operation threads of the ASP operator and the safety supervisor, respectively; the dashed and solid lines indicate the unexecuted and executed threads, respectively. The red dotted arrow line from the ASP operator icon to the operation advisor icon indicates the transfer of control authority to the operation advisor from the ASP operator and vice versa. $A_4$ and $D_4$ represent the start and end states of the DLC task, respectively; $B_4$ and $B_4^{'}$ are the same state and represent the ASP operator hands over the control authority to the safety supervisor who takes over the operation; $C_4$ and $C_4^{'}$ are the same state and represent the safety supervisor completes its operations, and the ASP operator continues the operation. $B_4^{'} \to C_4$ represents the operation of the safety supervisor, i.e., to take over the operation. At state $C_4^{'}$, the ASP operator issues the command for manual operation, and the safety supervisor returns the control authority to the ASP operator, who continues the operation based on the state $C_4^{'}$ ($C_4$).

Denote the time steps corresponding to state $B_4$ ($B_4^{'}$) and $C_4$ ($C_4^{'}$) are $t_4$ and $t_4+l_4$, the pseudocode of the safety supervisor is shown in **Algorithm 5**.

**Algorithm 5.** The safety supervisor.
1: **Step 1:** Read the process state of $B_4$.
2: **Step 2:** Stop the timer of the ASP operator, and start the timer of the safety supervisor.
3: **for** every time step $t = t_4, t_4+1, \cdots, t_4+l_4$ **do**
4:   **Step 3:** Call $IMPC(t)$ and calculate $\Delta U^*(t)$:
5:     $U^*(t) = IMPC(t)$.
6: **end for**
7: **Step 4:** Stop the timer of the supervisor, and start the timer of the ASP operator.

*5. The SO acts as the troublemaker.*

The function of the troublemaker is to deliberately make unreasonable operational behaviors that may occur in the physical process. The troublemaker aims to cultivate the ASP operator's ability to evaluate and analyze the behavior of the machine algorithm (the nonlinear IMPC). The troublemaker is designed to make three types of trouble: incorrect SSO targets (type I trouble), uncoordinated dynamic control actions (type II trouble), and inaccurate process measurements of the CVs (type III trouble).

Typically, a mismatched prediction model can result in inaccurate SSO targets. The model mismatch can occur if the prediction model is not effectively maintained during the long-time operation. Therefore, the first trouble can be reproduced by modifying the LP problem in Eq. (8) as follows:

$$\left(U_{sso}^{\times}(t), Y_{sso}^{\times}(t)\right) = \underset{U_{sso}(t), Y_{sso}(t)}{\arg \min} J_{SSO}(t) = $$
$$B^T U_{sso}(t) + C^T Y_{sso}(t) + Z^T S(t)$$
$$s.t.$$
$$Y_{sso}(t) = K_M^{\times}(w) U_{sso}(t) + D(t) \quad (21)$$
$$u_{i,\min} \leq u_i(t) \leq u_{i,\max}$$
$$y_{k,\min} - s_k(t) \leq y_k(t) \leq y_{k,\max} + s_k(t)$$
$$i = 1, \cdots, n_u, k = 1, \cdots, n_y$$

where $U_{sso}^{\times}(t)$ and $Y_{sso}^{\times}(t)$ are the wrong optimization results caused by model mismatch; $K_M^{\times}(w)$ is the wrong gain matrix of the mismatch model; the other variables are identical to those in Eq. (8). It is dangerous for the ASP to send the wrong tracking targets to the NDPC module, which will cause the non-conservation of materials and destroy the stability of process working conditions. Note that the model mismatch will also affect the performance of the NDPC module because the prediction accuracy of the prediction model will deteriorate.

The second type of trouble is attributed to the uncoordinated control rate of the multiple MVs in the NDPC module. If the control rates of multiple MVs are not coordinated, the working conditions of the ASP will significantly fluctuate. Typically, this trouble occurs during the design phase of the machine algorithm or after equipment maintenance due to poor knowledge of the equipment characteristics, such as specific valve characteristics. This trouble can be reproduced by modifying the lower and upper limits of the MV increments for the QP problem in Eq. (15) as follows:

$$U_M^{\times,iter}(t) = \arg\min_{U_M(t)} J_{NDPC}(t) =$$

$$\sum_{l=1}^{P}\left\|Y^{ref}(t+l) - Y_{PM}(t+l|t)\right\|_{Q_l}^2 + \sum_{l=1}^{P}\left\|\mathrm{E}(l)\right\|_{H_l}^2$$

$$+\sum_{l=0}^{M-1}\left\|\Delta U(t+l)\right\|_{R_l}^2 + \sum_{l=0}^{M-1}\left\|U(t+l) - U_{sso}^*(t)\right\|_{V_l}^2$$

s.t.

Prediction equation constraints : Eqs.(11−13)

$$y_k^{ref}(t+l_y) = y_k(t) + \left(y_{k,sso}^*(t) - y_k(t)\right)\left(1 - e^{-l_y T/\tau_k}\right) \quad (22)$$

$$y_{k,\min} - \varepsilon_k(t+l_y) \le y_{k,PM}(t+l_y|t) \le y_{k,\max} + \varepsilon_k(t+l_y)$$

$$\Delta u_i(t+l_u) = u_i(t+l_u) - u_i(t+l_u - 1)$$

$$u_{i,\min} \le u_i(t+l_u) \le u_{i,\max}$$

$$\Delta u_{i,\min}^{\times} \le \Delta u_i(t+l_u) \le \Delta u_{i,\max}^{\times}$$

$$l_y = 1,\cdots,P, l_u = 0,\cdots,M-1$$

$$k = 1,\cdots,n_y, i = 1,\cdots,n_u$$

where $\Delta u_{\min}^{\times}$ and $\Delta u_{\max}^{\times}$ are the deliberately modified lower and upper limits of the control movement increments; $U_M^{\times,iter}(t)$ is the uncoordinated action for the physical process, and the other variables are identical to those in Eq. (15).

The inaccurate process measurements of CVs cause the last type of trouble. The performance of the machine algorithm (the nonlinear IMPC) relies on reliable instrument measurements. Engineers must regularly calibrate the measuring instruments to ensure that the algorithm makes the correct decisions in industrial applications. If the measuring instrument fails (for example, zero drift), the measured value will deviate from the real value, forcing the algorithm to make wrong decisions. The serious nitrogen blockage accident made by the IMPC shown in Fig. 2 has emphasized the importance of high-precision measurements. This trouble can be reproduced by artificially causing the deviation between the measurement value required by the algorithm and the actual value. The optimization problem in Eq. (15) is modified as follows:

$$U_M^{\times,iter}(t) = \min_{U_M(t)} J_{NDPC}(t) =$$

$$\sum_{l=1}^{P}\left\|Y^{ref}(t+l) - Y_{PM}(t+l|t)\right\|_{Q_l}^2 + \sum_{l=1}^{P}\left\|\mathrm{E}(l)\right\|_{H_l}^2$$

$$+\sum_{l=0}^{M-1}\left\|\Delta U(t+l)\right\|_{R_l}^2 + \sum_{l=0}^{M-1}\left\|U(t+l) - U_{sso}^*(t)\right\|_{V_l}^2$$

s.t.

Prediction equation constraints : Eqs.(11−13)

$$y_k^{ref}(t+l_y) = y_k^{\times}(t) + \left(y_{k,sso}^*(t) - y_k^{\times}(t)\right)\left(1 - e^{-l_y T/\tau_k}\right) \quad (23)$$

$$y_{k,\min} - \varepsilon_k(t+l_y) \le y_{k,PM}(t+l_y|t) \le y_{k,\max} + \varepsilon_k(t+l_y)$$

$$\Delta u_i(t+l_u) = u_i(t+l_u) - u_i(t+l_u - 1)$$

$$u_{i,\min} \le u_i(t+l_u) \le u_{i,\max}$$

$$\Delta u_{i,\min} \le \Delta u_i(t+l_u) \le \Delta u_{i,\max}$$

$$l_y = 1,\cdots,P, l_u = 0,\cdots,M-1$$

$$k = 1,\cdots,n_y, i = 1,\cdots,n_u$$

where $y_k^{\times}(t)$ is the modified measurement value, $U_M^{\times,iter}(t)$ is the wrong action for the physical process, and the other variables are identical to those in Eq. (15).

Setting the above three types of trouble (other troubles, such as combinations of three troubles, can be designed) helps the ASP operator improve the ability to evaluate and analyze the behavior of the machine algorithm. Furthermore, the ability of operators to deal with accidents caused by the machine algorithm can be enhanced.

To summarize Section 3, the five identities, functions, and algorithms of the virtual instructor (SO) that interacts with and trains the ASP operators during the DLC operation have been designed.

## 4. Industrial application

This section first presents the application of the proposed five SO identities in DLC tasks and then discusses the effect of the SO on operational skills training.

### 4.1. Application of the five SO identities in DLC tasks

The nonlinear IMPC machine algorithm introduced in Section 3.1 is the basis of the SO. The LPV prediction model in Eq. (5) is obtained by the identification method in [6]. Specifically, the asymptotic method [22] is used to identify the local models at steady-state working points, and the weighting functions are the cubic spline function [6]. The LP and QP problems in Eqs. (8) and (15) are implemented using CasADi [23] and solved using IPOPT [24]. The control period of the IMPC is 0.5 min. Table 2 lists the MVs and part of the CVs of the nonlinear IMPC algorithm.

Fig. 3 shows that the ASP has four steady-state working points: 18,000, 19,000, 20,000, and 21,000 Nm$^3$/h. Therefore, 12 DLC tasks are designed [6]. The SO can interact with the ASP operator in DLC tasks between any two different working points. Here, the application results of the five SO identities are presented by taking the 18,000–19,000 Nm$^3$/h DLC task as an example. Specifically, the GOX load (CV2) needs to be adjusted from 18,000 to 19,000 Nm³/h, and all safety and purity constraints must be met.

Table 2. MVs and partial CVs of the nonlinear IMPC algorithm.

| Symbol | Tag | Description | Unit |
| --- | --- | --- | --- |
| MV1 | CCSSV_Q | Set value of feed air | Nm³/h |
| MV2 | HIC102 | Flow control valve position of the product GOX | % |
| MV3 | FIC103 | Flow control valve position of product GAN | % |
| MV4 | HIC3 | Flow control valve position of pure LIN of HPC | % |
| MV5 | HIC705 | Flow control valve position of crude GAR | % |
| MV6 | PIC104 | Control valve position of waste nitrogen pressure | % |
| MV7 | PICS_3302 | Control valve position of oxygen compressor | % |
| MV8 | LIC701 | Control valve position of liquid condenser level of CAC-I | % |
| MV9 | HC8 | Flow control valve position of product LIN | % |
| MV10 | FIC1 | Flow control valve position of bypass air | Nm³/h |
| CV1 | FI101 | Total air feed flow | Nm³/h |
| CV2 | FI102 | Product GOX flow (GOX load) | Nm³/h |
| CV3 | FI103 | Product GAN flow | Nm³/h |
| CV4 | AI701 | Argon content at the feed of CAC-I | % |
| CV5 | AIAS102 | Oxygen content of product GOX | % |
| CV6 | AIAS103 | Oxygen content of product GAN | ppm |
| CV7 | AI705 | Argon content at the top of CAC-II | % |

*1. The application result of the task performer.*

The task performer aims to complete and demonstrate the specific DLC task. AI701 (CV4) is selected as the key indicator to show the fluctuations in the ASP. Fig. 9 shows the operation result of the task performer.

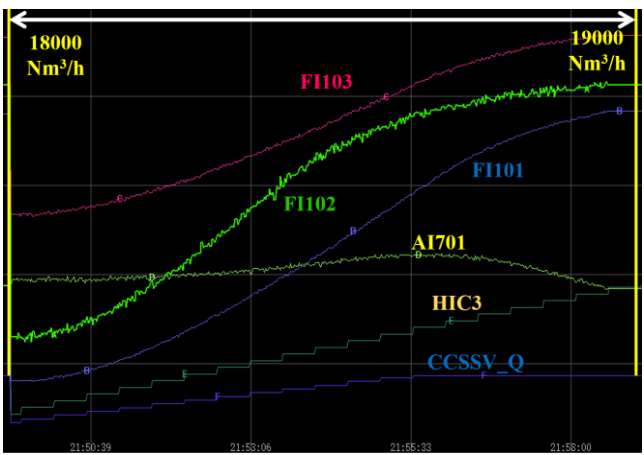

Fig. 9. Operation result of the task performer.

As shown in Fig. 9, the CCSSV_Q (MV1) and HIC3 (MV4) rise in steps every 0.5 min by solving the LP and QP problems in Eqs. (8) and (15). The operation shown in this figure is excellent because the key materials FI101 (CV1), FI102 (CV2), and FI103 (CV3) rise in a coordinated manner, and the entire operation takes only 12 min with very small process fluctuations (see the AI701 curve). The operation indicates that the IMPC machine algorithm in Section 3.1 is effective, and the SO is qualified as a training teacher in terms of operational skills.

*2. The application result of the operation advisor.*

The cooperation partner aims to work with the ASP operator to complete the specific DLC task. The ASP operator can determine the control authority of any MV at any time. Fig. 10 shows the user interface (UI) of the cooperation partner and an application scenario.

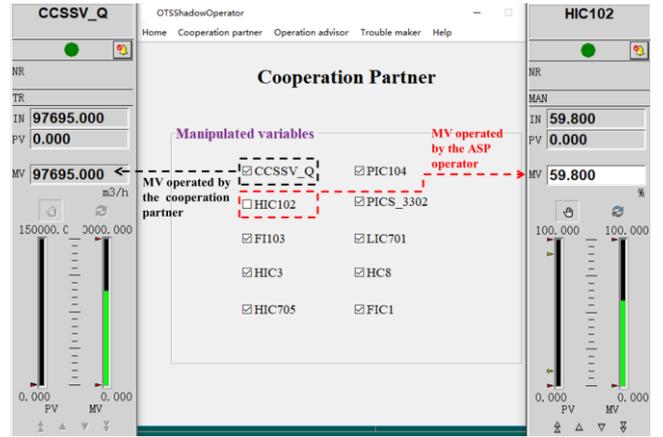

Fig. 10. UI of the cooperation partner.

In Fig. 10, all MVs listed in Table 2 are shown in the UI; ☑ and ☐ denote the MVs operated by the partner and ASP operator, respectively. The figure depicts an application scenario, with the HIC102 (MV2) in the red dotted box indicating that the ASP operator controls the MV while the cooperation partner operates the rest. A novice ASP operator was invited to perform the DLC task with the cooperation partner, and the operation result is shown in Fig. 11.

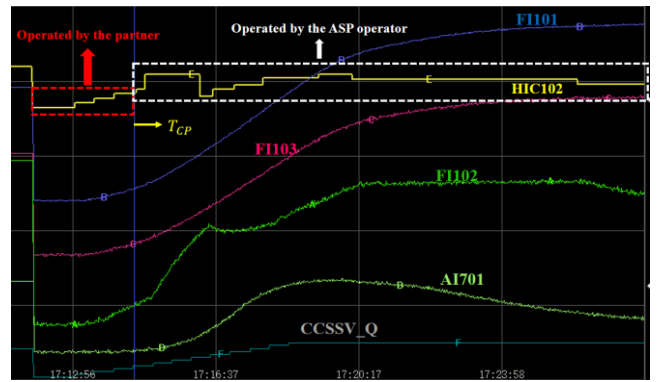

Fig. 11. Operation result of the cooperation partner.

As shown in Fig. 11, the cooperation partner operated all MVs before the time step $T_{CP}$. At $T_{CP}$, the ASP operator operated the HIC102 manually, and the cooperation partner operated the rest MVs (see the scenario in Fig. 10). However, due to the bad operation of the ASP operator, FI102 did not rise in coordination with other flow variables (FI101 and FI103), causing the AI701 to fluctuate. To master the complex operational skills required to operate the 10 MVs in Table 2 simultaneously, the ASP operator requires extensive and effective training. The cooperation partner works as a teammate and shares operational responsibilities during training.

*3. The application result of the operation advisor.*

The operation advisor aims to provide operational advice when the operator issues a help request. During the operation, the operator must operate all MVs independently, and the advisor will advise when it is activated. Fig. 12 shows an application scenario in which the operation advisor provides one-step-ahead operational advice.

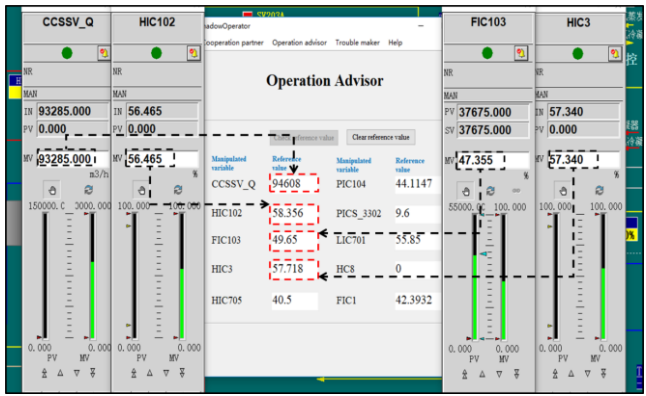

**Fig. 12.** One-step-ahead operational advice given by the operation advisor.

In Fig. 12, the values in the black dashed boxes are the current operation values (only four MVs are shown), and the values in the red dashed boxes are the recommended values for the next manual operation. All recommended MV values for the next operation are shown in the UI, and the ASP operator can adjust operation behavior based on the recommendations.

*4. The application result of the safety supervisor.*

The safety supervisor aims to take over the operation when the operator transfers the control authority to the supervisor. Fig. 13 shows the operation result of the safety supervisor.

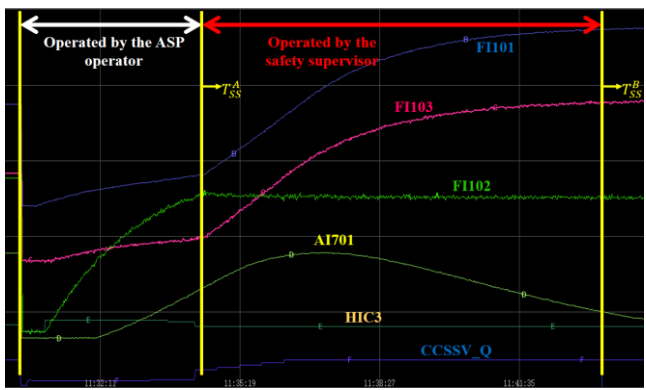

**Fig. 13.** Operation result of the safety supervisor.

As shown in Fig.13, the operator significantly adjusted the FI102 to the target during the initial phase. However, the AI701 rose at a high rate, and large process fluctuations occurred because the other materials did not coordinate with FI102 (the rise of FI102 is considerably faster than that of FI101 and FI103). A high AI701 value is one of the causes of a nitrogen blockage accident. At the time step $T_{SS}^A$, the operator delegated the control authority to the safety supervisor. Although the AI701 curve rose after the supervisor took over the operation due to the slow process dynamics and hysteresis characteristics, the safety supervisor eventually restored the working condition to normal. At $T_{SS}^B$, the supervisor returned the control authority to the operator.

*5. The application result of the troublemaker.*

The troublemaker aims to cultivate the ASP operator's ability to analyze and evaluate the behavior of the machine algorithm (the IMPC). The troublemaker is set to have ten operation modes, including nine abnormal operation modes (making various incorrect decisions) and one normal operation mode (making the correct decisions), where the operation mode is chosen at random. Fig. 14 shows the UI of the troublemaker.

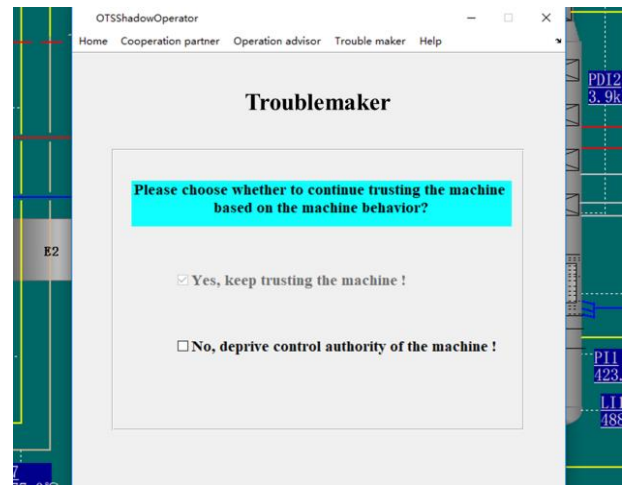

**Fig. 14.** UI of the troublemaker.

Fig. 14 shows two options presented to the ASP operator. The first and default option is to keep trusting the machine algorithm. The operator requires to decide whether to switch to the second option (deprive control authority of the machine) based on the machine behavior. Fig. 15 shows a trouble case made by the troublemaker.

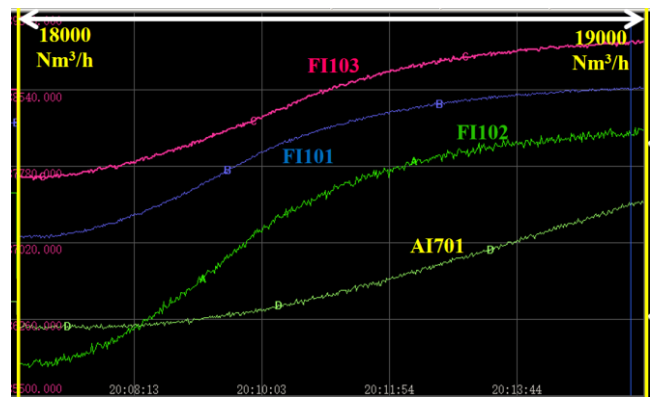

**Fig. 15.** A trouble case made by the troublemaker.

As shown in Fig. 15, FI102 was normally adjusted from 18,000 to 19,000 Nm$^3$/h. However, due to the incorrect SSO targets and uncoordinated control actions (type I and type II troubles are combined in this case), FI101 and FI103 failed to coordinately adjust to the corresponding targets, causing significant process fluctuations (see AI701 in the figure. The ASP operator must quickly select the second option in Fig. 14 before large fluctuations occur (at least before the DCS safety alarm system is triggered) and control the ASP to a safe state as soon as possible.

So far, the application of the proposed five SO identities in the DLC task has been presented. This research then applies the SO to operation training and investigates its effects.

### 4.2. Effect of the SO on operational skills training

In order to verify the effectiveness of SO on operational skills training compared to traditional training methods in which the human operator acts as an instructor, this section designs human factors experiments, described as follows.

#### 4.2.1. Experimental protocol

First, the experimental and control groups are set up. The SO, as a virtual instructor, instructs the experimental group participant (the proposed human–machine interactive training method). A skilled ASP operator instructs the control group participant (the traditional training method). The skilled operator in the control group is the ASP chief operator, who scored an average of 97.6 (out of a possible 100) points in the May 2021 Load Change Operations Career Skills Competition [6].

Then, the following two comparative experiments are designed:
(1) Experiment 1 compares the DLC operational skill of participants in the experimental and control groups, including the skill learning curves and the stable skill scores after numerous training exercises. The task performer, cooperation partner, operation advisor, and safety supervisor instruct experimental group participants. The troublemaker identity of the SO is not activated in Experiment 1.
(2) Experiment 2 compares the abilities of participants in the experimental and control groups when the machine algorithm (the troublemaker) makes abnormal decisions, such as comprehending the machine behavior and responding appropriately to abnormal situations made by the algorithm.

**Remark 1.** Each operation will be quantified by the skill evaluation algorithm in [6] as a score from 0 to 100 points. The evaluation algorithm evaluates the operational skills with four indicators: plant safety, product purity, task completion time, and energy consumption.

After a DLC skill examination, two participants with comparable initial skill levels were selected from five novice operators of similar ages and randomly assigned to the experimental and control groups. The two participants had never performed DLC operations on the physical ASP and had attempted fewer than four times to use the deployed OTS.

Moreover, the two operators have mastered feedback control and proportional integral derivative control techniques, as well as the equipment safety and operation principles of the ASP.

#### 4.2.2. Experimental results and analysis

1. Experiment 1.

In the first experiment, the participants in the experimental and control groups were subjected to periodic examinations to track their skill learning process. During the examination, the two participants must manually complete the temporarily assigned DLC task without assistance. Fig. 16 shows the skill learning curves of the two participants after 27 examinations. The two curves were obtained by fitting a fifth-order polynomial to two score datasets.

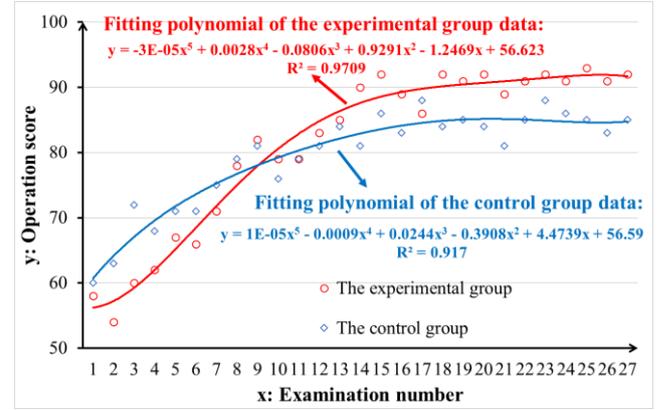

**Fig. 16.** Learning curves of the two participants.

In Fig. 16, the red circle and blue diamond represent the operation scores of the experimental and control operators, respectively; the red and blue lines represent the learning curves of the experimental and control operators, respectively. The fitting polynomials for the two groups are given as follows:

$$y^E = (-3E-05)x^5 + 0.0028x^4 - 0.0806x^3 \\ +0.9291x^2 - 1.2469x + 56.623 \quad (24)$$

$$y^C = (1E-5)x^5 - 0.0009x^4 + 0.0244x^3 \\ -0.3908x^2 + 4.4739x + 56.59 \quad (25)$$

where $y^E$ denotes the experimental group and $y^C$ denotes the control group.

Fig. 16 is analyzed as follows:
(1) The apparent upward trend in the scores of the two operators with increasing training times indicates that both training methods can improve the DLC skills of the two operators.
(2) In the first eight examinations, the operator in the experimental group scored lower than the operator in the control group. This may have occurred because the experimental group operator was unfamiliar with the behavior of the SO during early training. From the ninth examination, the experimental group operator scored higher than the control group operator. The rising rate of the learning curve in the experimental group was significantly higher than that in the control group, indicating that the SO accelerated the process of the DLC skill learning for the experimental group operator.

(3) From the 22nd examination, the two operators obtained stable scores. The average scores of the experimental and control group operators in the last six examinations were 91.7 and 85.3 points, respectively, indicating that the SO allowed the experimental group operator to reach a higher stable skill level.

2. *Experiment 2.*

In Experiment 2, the SO (the troublemaker) was programmed to make different wrong decisions in three DLC tasks (18,000–19,000, 18,000–19,000, and 18,000–20,000 $Nm^3/h$) at specific times. The two operators must deprive the control authority of the SO quickly (see Fig. 14) and adjust the working conditions to the target. The time for the SO to make wrong decisions, the time for the operator to operate manually and the final score for the operation were recorded. Figs. 17 and 18 show the results of Experiment 2.

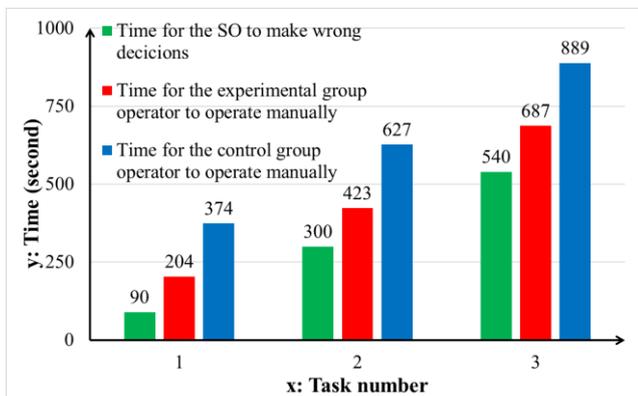

**Fig. 17.** Time (in seconds) for the SO to make wrong decisions and for the two operators to manually operate in three DLC tasks.

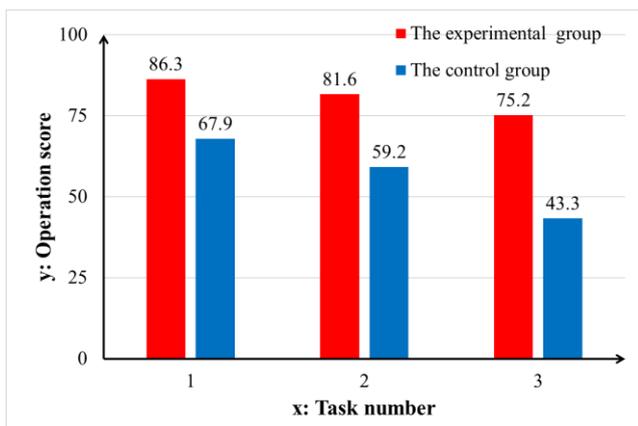

**Fig. 18.** Scores for the two operators in three DLC tasks.

Figs. 17 and 18 are analyzed as follows:
(1) In Fig. 17, the time for the two operators to take over the operation lagged far behind the time for the SO to make wrong decisions, and the minimum lag time was 114 seconds. This is due to the slow dynamics and hysteresis of the ASP. The impact of a wrong decision takes a while to manifest.

(2) In Fig. 17, the operator from the experimental group took over the operation significantly earlier than the operator from the control group. This can be explained by the fact that the operator trained by the SO is more familiar with the behavior and thus can more quickly identify the wrong decisions (strange behavior) made by the SO and take back the control authority.

(3) In Fig. 18, the scores for the control group operator were significantly lower than those in the experimental group. Combined with Fig. 17, it can be explained that the control group operator took over the operation later, and the initial condition faced by the control group operator was worse than that of the experimental group operator. The control group operator requires more time to restore the working conditions and complete the DLC task, and the evaluation algorithm takes operation time into account [6]. Moreover, Experiment 1 shows that the control group operator had lower stable DLC operational skills than the experimental group operator, which affects the score in Experiment 2.

From the experimental results and analysis, it can be concluded that the SO enables novice ASP operators to achieve higher DLC operational skills more quickly and helps operators effectively identify the wrong decisions made by the machine algorithm, thereby improving the safety of dynamic operations.

**5. Conclusions**

Considering that machine algorithms are deeply involved in the daily DLC operations of the ASP, a novel human–machine interactive training method is proposed. A SO is developed as a virtual instructor to train ASP operators in this method. First, a nonlinear IMPC machine algorithm is developed. The IMPC employs an LPV model and executes precise operations using a multi-step linearization algorithm. Second, to improve the effectiveness of DLC operation training, this study establishes an HMC model. The model is inspired by an educational psychology framework (Bloom's taxonomy) and facilitates the ASP operators' skill acquisition and improvement. Finally, five SO identities are designed to offer various dynamic training modes for ASP operators during DLC operations. The SO is used to train novice operators, and the application demonstrates its beneficial contributions to operation training.

Future work will focus on the following aspects:
(1) The accident shown in Fig. 2 indicates that the IMPC machine algorithm must be improved to be more fault-tolerant when actual measurements are biased. Methods for sensor faults estimation [25] and fault-tolerant control [26] are being studied.
(2) The results in Section 4.2 can be strengthened by more experiments. More participants will be recruited, and the impact of gender, age, and trust in machine behavior [27] on skill training will be examined.

**Acknowledgments**

This study is financially supported by the National Natural Science Foundation of China (Nos. 62173301 and 62120106003) and the Key Research and Development Program of Zhejiang Province (No. 2021C01151). The authors are grateful to the operators of Gas Supply Company of Nanjing Iron Steel United Co., Ltd. for meaningful discussions and active cooperation.